\documentclass[aps,pra,groupedaddress,reprint,floatfix,notitlepage,balancelastpage,showpacs]{revtex4-1}
\usepackage{graphicx,amsmath,amssymb,bm,color,engord}
\usepackage[usenames,dvipsnames]{xcolor}
\usepackage[colorlinks,colorlinks,urlcolor=Magenta,citecolor=Magenta,linkcolor=black]{hyperref}
\begin{document}
\title{Hybrid of superconducting quantum interference device and atomic Bose-Einstein condensate: An architecture  for  quantum information processing}
\author{Kelly R. Patton and Uwe R. Fischer }
\affiliation{Seoul National University, Department of Physics and Astronomy\\ Center for Theoretical Physics, 151-747 Seoul, Korea }
\date{\today}
\begin{abstract}
A  hybrid quantum system is proposed by coupling  the internal hyperfine transitions of a trapped  atomic Bose-Einstein condensate (BEC)  and a superconducting quantum interference device (SQUID) via the macroscopic quantum field of the flux qubit.  The presence of the condensate leads to a bosonic enhancement of the Rabi frequency over the otherwise small single-particle magnetic dipole transition matrix elements.  This enhancement allows for the possibility to rapidly transfer and store qubit states in the BEC that were originally  prepared in the SQUID.    The fidelity of this transfer for different states is calculated, and a direct experimental protocol to determine the transfer fidelity by quantum tomography of the BEC qubit is presented. 
\end{abstract}
\maketitle
\section{Introduction}
Hybrid quantum systems, systems composed of two or more distinct  quantum mechanical subsystems coupled together,  is a rapidly developing  area, and a subject that intrinsically spans many disciplines, from basic research to engineering \cite{Xiang,Aspelmeyer}.    On a fundamental level, such a combined system was used to probe---for the first time---the quantum nature of a  nanomechanical oscillator, by coupling it to a superconducting qubit \cite{OConnell}.   These systems have especially become prominent in the quantum computing community, where such combinations can, for example, be used to overcome the opposing  requirements for a  quantum computer to have both long coherence times and an ease of external manipulation \cite{Blais, Verdu, Lukin, Armour,Nakamura,Wallraff,Zhu,Wu, Majer,Cleland,Rabl,Brennecke,Baumann}.   Frequently, such hybrid systems are not developed as algorithmic units but to perform auxiliary functions, such as  memory elements, where qubit states are transferred from one subsystem and stored in another, or as information buses \cite{Majer,Cleland,Rabl,Brennecke,Baumann,Blencowe}.   

Here, we propose a novel quantum memory hybrid, where the storage element is  potentially capable of possessing second-long coherence times, competing with the recently engineered  coherence times of nitrogen-vacancy (NV) centers  in diamond, without the use of a dynamical environment-decoupling scheme \cite{vanderSar}.  This hybrid is created by coupling a SQUID, or flux qubit, and the hyperfine states of a trapped atomic Bose-Einstein condensate (BEC).  The SQUID allows for the easy preparation and manipulation of qubit states, while the BEC's  hyperfine qubit states can remain coherent for a extremely long time, on the order of seconds. The latter has been recently demonstrated for a BEC cloud trapped near a superconducting waveguide resonator geometry \cite{BernonArxiv2013}, further motivating our study into the present hybrid. 

The two qubits are electromagnetically coupled together.  The radiation emitted by the circulating macroscopic currents of the flux qubit induces atomic transitions of the trapped BEC atoms.   For the system considered here, the transitions are dominated by magnetic dipole coupling but, in general could also be  electric dipole driven.  The hyperfine transitions can in turn excite the qubit states of the SQUID, resulting in the periodic transfer of energy from one subsystem to the other, i.e., the Rabi process, with a period determined by a Rabi frequency $\Omega$.  
Qubit states initially prepared in the SQUID can be transferred and stored in the BEC. This process is achieved by dynamically bringing the two subsystems into and out of resonance for approximately a half of a Rabi period $\pi/(2\Omega)$.  

Present coherence times of flux qubits are on the order of a few microseconds. Depending on the BEC-SQUID separation, the qubit-state-transfer time can  be of the same order.  But as was recently shown in Ref.~\cite{Patton2012}, one can still obtain a high transfer fidelity, even in the face of this naively unfavorable condition.  

In the following sections each qubit subsystem will be introduced, starting with the SQUID in Sec. \ref{SQUID}, followed by the hyperfine qubit in Sec. \ref{qubit BEC}, and then the qubit-qubit coupled hybrid system is presented in Sec. \ref{Hybrid system}.  Finally, in Secs. \ref{state transfer fidelity} and \ref{BEC tomography} the transfer and storage fidelity is calculated for various qubit states, and an experimental technique is put forward to determine the hyperfine qubit density matrix by quantum tomography. 
\section{Flux qubit}
\label{SQUID}
Although the derivation of the effective low-energy Hilbert space of a SQUID  can be commonly found in the literature \cite{SQUIDHandbook}, starting from a quantum circuit model, which is used here, or from a microscopic approach \cite{Ambegaokar1982}, several details relevant to the BEC-SQUID coupling are typically not prominently discussed, such as defining a macroscopic current operator for the SQUID and the resulting quantum mechanical electromagnetic fields.   Therefore,  to make the paper sufficiently self-contained a short review is included here.    

For simplicity and clarity, in the following the SQUID is assumed to be a simple rf SQUID, containing a single Josephson tunneling junction, see Fig.~\ref{fig1}.  The outcome is not dependent on this, and the intermediate details can be readily generalized to other more complex SQUID architectures. 
\subsection{Quantum circuit model for a SQUID}
The supercurrent $I_{\rm s}$ through a weak-link tunneling junction is given by the DC Josephson relation, 
\begin{equation}
\label{dc Josephson effect}
I_{\rm s}=I_{\rm c}\sin \delta,
\end{equation}
where $I_{\rm c}$ is the critical current of the junction, which depends on the microscopic details of the barrier, and $\delta$ is the gauge-invariant phase difference of the superconducting states on each side of the tunneling barrier.  Additionally, the voltage across the tunnel junction is related to the time-rate change of the phase difference  by Josephson's second relation
\begin{equation}
\label{ac Josephson effect}
V(t)=\frac{\hbar}{2\rm {e}}\frac{ d\delta}{dt}=\frac{\Phi^{}_{0}}{2\pi }\frac{d\delta}{dt},
\end{equation}
where $\Phi^{}_{0}=\frac{h}{2\rm {e}}$ is the quantum of magnetic flux. 
The total flux $\Phi$ through the SQUID loop and  the phase are related by the flux quantization condition 
\begin{equation}
\Phi+\Phi^{}_{0}\delta/(2\pi)=n\Phi^{}_{0},
\end{equation}
with $n\in \mathbb{Z}$.  The Josephson relations, Eqs.~\eqref{dc Josephson effect} and  \eqref{ac Josephson effect}, in terms of the flux variable are 
\begin{subequations}
\label{Josephson relations flux}
\begin{align}
I_{\rm s}&=-I_{\rm c}\sin\left(\frac{2\pi}{\Phi^{}_{0}}\Phi\right),\label{Josephson current-flux relation}\\
V&=-\frac{d\Phi}{dt}\label{Josephson voltage-flux relation}.
\end{align}
\end{subequations}
The total flux through the loop is  related to the total current $I$ by
\begin{equation}
\label{flux relation}
\Phi=\Phi^{}_{\rm ex}+LI,
\end{equation}
where $\Phi^{}_{\rm ex}$ is any applied external flux and $L$ is the geometric inductance of the SQUID.  

\begin{figure}
\includegraphics[width=0.5\columnwidth]{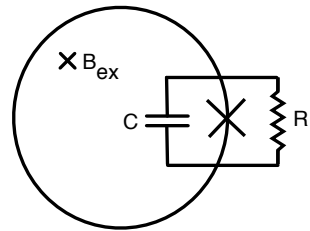}%
\caption{Schematic of a rf SQUID in an external magnetic field $B_{\rm ex}$. A realistic Josephson junction is modeled by an ideal junction, denoted by the cross, in parallel with a capacitor $C$ and a resistor $R$.   \label{fig1}}
\end{figure}

Using the  current-voltage relations for a capacitor $C$, $I=C\partial_{t}V$, Ohm's law $I=R^{-1}V$, for a resistor $R$, and  Eqs.~\eqref{Josephson relations flux}, the time-dependent current around the loop of the rf SQUID model shown in Fig.~\ref{fig1} is given by

\begin{equation}
\label{total SQUID current}
I(t)=-I_{\rm c}\sin\left(\frac{2\pi}{\Phi^{}_{0}}\Phi\right)-C\frac{d^{2}\Phi}{dt^{2}}-\frac{1}{R}\frac{d\Phi}{dt}.
\end{equation}
 
Putting the current  \eqref{total SQUID current} into  \eqref{flux relation} leads to an equation of motion for the flux variable given by
\begin{equation}
\label{classical equation of motion}
C\frac{d^{2}\Phi}{ dt^{2}}+\frac{1}{R}\frac{d\Phi}{dt}=-\partial^{}_{\Phi}U(\Phi)
\end{equation}
where
\begin{equation}
\label{potential function}
U(\Phi)=\frac{\left(\Phi^{}_{}-\Phi^{}_{\rm ex}\right)^{2}}{2L}-\frac{I_{\rm c}\Phi^{}_{0}}{2\pi}\cos\left(\frac{2\pi}{\Phi^{}_{0}}\Phi\right). 
\end{equation}
Sometimes it is convenient to express \eqref{potential function} using dimensionless quantities 
\begin{equation}
\label{potential in reduced units}
U(\phi)=U_{0}\left[\frac{(2\pi)^{2}\big(\phi-\phi_{\rm ex}\big)^{2}}{2}-\beta^{}_{L}\cos(2\pi\phi)\right],
\end{equation} 
where $U_{0}=\Phi^{2}_{0}/(4\pi^{2}L)$, $\beta^{}_{L}=2\pi LI_{\rm c}/\Phi^{}_{0}$ and $\phi$ and $\phi_{\rm ex}$ are in units of $\Phi^{}_{0}$.

Equation \eqref{classical equation of motion} is the  equation of motion  identical to that of a fictitious particle having a mass $C$ in the presence of a damping, or dissipation, $1/R$ term  and potential $U(\Phi)$.   Ignoring the dissipation term, valid in the $R\to \infty$ limit, a Lagrangian that produces  \eqref{classical equation of motion} as an equation of motion is
\begin{equation}
\label{Lagrangian}
{\cal L}=\frac{1}{2} C \dot{\Phi}^{2}-U(\Phi).
\end{equation} 
A Hamiltonian can be constructed from the Lagrangian in the usual way. Defining the flux $\Phi$ as the generalized coordinate, with conjugate momentum  $p^{}_{\Phi}=\partial_{\dot{\Phi}} {\cal L}=C\dot{\Phi}$,  the Hamiltonian  is 
\begin{align}
\label{Hamiltonian}
H&=p^{}_{\Phi}\dot{\Phi}-\cal{L}\nonumber\\&=\frac{p^{2}_{\Phi}}{2C}+U(\Phi). 
\end{align}
At the classical level, the Poisson bracket of the generalized coordinate $\Phi$ and its conjugate momentum $p_{\Phi}$ is unity. Thus, one can quantize the SQUID Hamiltonian \eqref{Hamiltonian} by imposing the standard canonical commutations relations, i.e., $p_{\Phi}\rightarrow -i\hbar \partial_{\Phi}$ such that $[\Phi, p^{}_{\Phi}]=i\hbar$.  The energy eigenstate of \eqref{Hamiltonian} $\psi_{n}$ are then functions of the flux variable   $\psi_{n}(\Phi)$.  In general these eigenstates  have to be found numerically, but in some parameter regimes  approximate solutions can be  constructed.  In the next section the approximate  qubit states of the SQUID  are given.

\subsection{Qubit basis}
\label{Qubit basis}

\begin{figure}
\includegraphics[width=\columnwidth]{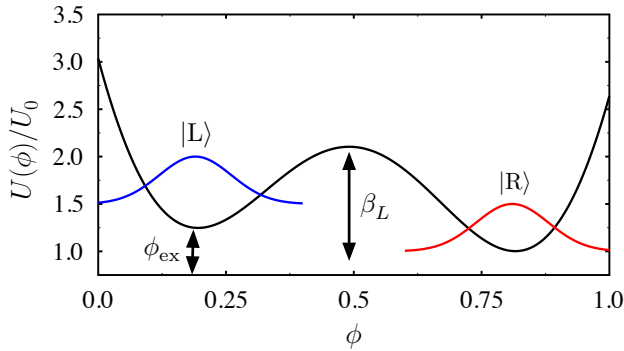}%
\caption{Characteristic potential felt by the fictitious  flux particle Eq.~\eqref{potential in reduced units}, where all flux values are in units of the flux quantum, i.e., $\phi=\Phi/\Phi^{}_{0}$, $\beta^{}_{L}=2.1$, and $\phi^{}_{\rm ex}=1.02/2$.  The location of the barrier and the asymmetry between the wells is controlled by the external flux $\Phi^{}_{\rm ex}$, while the barrier height is controlled by $\beta^{}_{\rm L}$. For $\beta^{}_{\rm L}=1$ the barrier vanishes.  Although in this setup $\beta^{}_{\rm L}$ is fixed, it can be dynamically controlled by introducing additional circuit elements. \label{fig2}}
\end{figure}
A typical plot of the characteristic double well of the effective potential \eqref{potential in reduced units} is shown in Fig.~\ref{fig2}. 
For a sufficiently high barrier that separates the two wells  one can approximate  the low-energy subspace of the Hilbert space, by  ground state harmonic oscillators, i.e., Gaussians,  centered in each well.  These will be labeled as $|{\rm L}\rangle$ and $|{\rm R}\rangle$ and are explicitly given by ($\hbar=1$ in what follows)  
\begin{equation}
\label{left and right states}
\langle\Phi|{\rm L}/{\rm R}\rangle=\left(\frac{C\omega^{}_{{\rm L}/{\rm R}}}{\pi}\right)^{1/4}\exp\left[-\frac{C\omega^{}_{{\rm L}/{\rm R}}}{2}\left(\Phi-\Phi^{\rm min}_{{\rm L}/{\rm R}}\right)^{2}\right],
\end{equation}
where $\Phi^{\rm min}_{{\rm L}/{\rm R}}$ is the location of the minimum of each well, and $\omega^{}_{{\rm L}/{\rm R}}$ is the effective harmonic  frequency of each well. These are related to the full potential  \eqref{potential function} by requiring $\partial^{}_{\Phi}U(\Phi^{\rm min}_{{\rm L}/{\rm R}})=0$, such that $\partial^{2}_{\Phi}U(\Phi^{\rm min}_{{\rm L}/{\rm R}})>0$, which implies 
\begin{equation}
\Phi^{\rm min}_{{\rm L}/{\rm R}}=\Phi^{}_{\rm ex}-\frac{\Phi^{}_{0}}{2\pi}\beta^{}_{L}\sin\left(\frac{2\pi}{\Phi^{}_{0}}\Phi^{\rm min}_{{\rm L}/{\rm R}}\right),
\end{equation}
 and 
\begin{align}
\omega^{}_{{\rm L}/{\rm R}}&=\sqrt{\partial^{2}_{\Phi}U(\Phi^{\rm min}_{{\rm L}/{\rm R}})/C}\nonumber\\&=\sqrt{\frac{1+\beta^{}_{L}\cos\left(\frac{2\pi}{\Phi^{}_{0}}\Phi^{\rm min}_{{\rm L}/{\rm R}}\right)}{LC}}\sim\sqrt{\frac{1}{LC}}. 
\end{align}
Because of the potential offset of each well, controlled by the external flux, see Fig.~\ref{fig2}, the ground state energy of each Gaussian is approximately $E_{{\rm L}/{\rm R}}=U(\Phi^{\rm min}_{{\rm L}/{\rm R}})+\omega^{}_{{\rm L}/{\rm R}}+\frac{1}{2}$.
For most cases of interest $E_{{\rm L}}-E_{{\rm R}}\approx U(\Phi^{\rm min}_{{\rm L}})-U(\Phi^{\rm min}_{{\rm R}}):=\varepsilon. $

The  left-right  states are not energy eigenstates. To see this, the SQUID Hamiltonian in this basis is 
\begin{align}
\label{left-right well state basis representation of the SQUID Hamiltonian}
\hat{H}&\approx\left(\begin{array}{cc}\langle{\rm L}|H|{\rm L}\rangle & \langle{\rm L}|H|{\rm R}\rangle \\ \langle{\rm R}|H|{\rm L}\rangle & \langle{\rm R}|H|{\rm R}\rangle\end{array}\right)\approx \left(\begin{array}{cc}E^{}_{\rm L} & -\Delta/2 \\ -\Delta/2 & E^{}_{\rm R}\end{array}\right)\nonumber\\&=\frac{E^{}_{\rm L} +E^{}_{\rm R} }{2}\openone+ \frac{\varepsilon}{2}\sigma_{z}-\frac{\Delta}{2}\sigma_{x},
\end{align}
where $\Delta$ characterizes the overlap of the left-right  states, or tunneling amplitude, and $\sigma_{i}$ are the Pauli matrices. The minus sign in front of the $\sigma_{x}$ term is chosen to give a symmetric lowest energy eigenstate, assuming $\Delta>0$. 
Finally, the term proportional to the identity matrix in \eqref{left-right well state basis representation of the SQUID Hamiltonian} can be dropped, as it's an over all constant, leading to the standard form of the flux qubit Hamiltonian,
\begin{equation}
\label{flux qubit Hamiltonian}
\hat{H}_{\rm S}=\frac{\varepsilon}{2}\sigma_{z}-\frac{\Delta}{2}\sigma_{x}. 
\end{equation} 
\subsection{Electromagnetic fields of the SQUID}
Although the $|\rm L\rangle$ and $|\rm R\rangle$ states are not energy eigenstates, they are (approximate) eigenstates of the current operator.  A current operator, as a function of the flux variable, can be defined from \eqref{flux relation} as, 
\begin{equation}
\label{current operator}
\hat{I}(\Phi)=\frac{\Phi-\Phi^{}_{\rm ex}}{L}.
\end{equation}
The eigenvalues of \eqref{current operator} correspond to  the experimentally measured current around the SQUID loop.
In the $|\rm L\rangle$ and $|\rm R\rangle$ basis 
\begin{align}
\hat{I}&=\left(\begin{array}{cc}\langle{\rm L}|I(\Phi)|{\rm L}\rangle & \langle{\rm L}|I(\Phi)|{\rm R}\rangle \\ \langle{\rm R}|I(\Phi)|{\rm L}\rangle & \langle{\rm R}|I(\Phi)|{\rm R}\rangle\end{array}\right)\nonumber\\&\approx\frac{1}{L}\left(\begin{array}{cc}\Phi^{\rm min}_{\rm L}-\Phi^{}_{\rm ex} & 0 \\0 & \Phi^{\rm min}_{\rm R}-\Phi^{}_{\rm ex}\end{array}\right).
\end{align}
For approximately symmetric wells $ \Phi^{\rm min}_{\rm L}-\Phi^{}_{\rm ex}\approx-(\Phi^{\rm min}_{\rm R}-\Phi^{}_{\rm ex}),$ thus
 \begin{equation}
 \label{Qubit space current operator}
 \hat{I}=I\sigma_{z},
 \end{equation}
 where $I=\langle{\rm L}|I(\Phi)|{\rm L}\rangle$.  Therefore, given that the ground state and first excited energy eigenstates of \eqref{flux qubit Hamiltonian} are linear combinations of $|\rm L\rangle$ and $|\rm R\rangle$, this implies coherent superpositions of left and right moving macroscopic current-carrying states span the low-energy Hilbert space of the SQUID.  These currents must also produce macroscopic electromagnetic fields.

Classically, in terms of vector ${\bm A}$ and scalar $\phi$ potentials the magnetic and electric fields are determined by
\begin{subequations}
\label{magnetic and electric field}
\begin{align}
{\bm B}({\bm r},t)&=\nabla\times {\bm A}({\bm r},t),\label{magnetic field}\\
{\bm E}({\bm r},t)&=-\nabla\phi({\bm r},t)-\partial^{}_{t}{\bm A}({\bm r},t)\label{electric field}.
\end{align}
\end{subequations}
In the Coulomb gauge the potentials can be expressed in terms of the charge  $\rho({\bm r},t)$ and current ${\bm j}({\bm r},t)$ densities by
\begin{subequations}
\label{scalar and vector potentials}
\begin{align}
\phi({\bm r},t)&=\frac{1}{4\pi \varepsilon_{0}}\int  d^{3}r'\, \frac{\rho({\bm r}',t)}{|{\bm r}-{\bm r}'|}\label{scalar potential}\\
{\bm A}({\bm r},t)&=\frac{\mu^{}_{0}}{4\pi}\int d^{3}r'\, \frac{{\bm j}({\bm r}',t)}{|{\bm r}-{\bm r}'|}\label{vector potential},
\end{align}
\end{subequations}
where $\varepsilon_{0}$ is the vacuum permittivity, and $\mu^{}_{0}$  the vacuum permeability. 
Neglecting  surface charges that form on the SQUID's circuitry,
\begin{subequations}
\label{approximate B and E}
\begin{align}
{\bm B}({\bm r},t)&=\nabla\times {\bm A}({\bm r},t)\label{approximate B}\\
{\bm E}({\bm r},t)&\approx-\partial^{}_{t}{\bm A}({\bm r},t)\label{approximate E}. 
\end{align}
\end{subequations}
Modeling the SQUID within classical electrodynamics as  a current-carrying loop of wire having radius $d$ with cross sectional radius $a$, such that $a\ll d$, the current density in spherical coordinates can then be approximated by
\begin{equation}
\label{SQUID current density}
{\bm j}({\bm r})=I\frac{\delta(r-d)}{d}\delta(\cos\theta)\hat{\bm e}_{\phi},
\end{equation}
where $I$ is the total current around the loop and $\delta(x)$ is a Dirac delta function.  Putting \eqref{SQUID current density} into \eqref{vector potential}  the vector potential can be expressed in closed form as
\begin{align}
\label{closed form for the vector potential}
{\bm A}({\bm r})&=\frac{\mu_{0}}{4\pi}\frac{4Id}{\sqrt{d^{2}+r^{2}+2dr\sin\theta}}\nonumber\\&\times\left[\frac{(2-k)K(k)-2E(k)}{k}\right]\hat{\bm e}_{\phi},
\end{align}
where $K(k)=\int_{0}^{\pi/2} dx\,(1-k^{2}\sin^{2} x)^{-1/2}$ and $E(k)=\int_{0}^{\pi/2} dx\,\sqrt{1-k^{2}\sin^{2} x}$ are the first and second complete elliptical integrals respectively, 
with $k=4dr\sin\theta/(d^{2}+r^{2}+2dr\sin\theta)^{-1}$ \cite{JacksonEM}.

The quantum mechanical vector potential  $\hat{\bm A}({\bm r})$ can then be defined by replacing the value of the current $I$ in \eqref{closed form for the vector potential} with the current operator $\hat{I}=I\sigma_{z}$ \eqref{Qubit space current operator} giving 
\begin{equation}
\hat{\bm A}({\bm r})={\bm A}(\bm r)\sigma_{z}.
\end{equation} 
The magnetic field operator is then
\begin{equation}
\label{magnetic field operator}
 \hat{\bm B}({\bm r})={\bm B}(\bm r)\sigma_{z},
 \end{equation}
  where ${\bm B}(\bm r)=\nabla\times {\bm A}({\bm r})$.  The electric field operator can be found by Heisenberg's equation motion
\begin{align}
\label{electric field operator}
\hat{\bm E}({\bm r})&=-\partial_{t} \hat{{\bm A}}(\bm r)=-i\big[\hat{H}_{\rm S},\hat{{\bm A}}({\bm r})\big]\nonumber\\&={\bm A}({\bm r})\Delta \sigma_{y}.
\end{align}
Note that operator notation in the present context refers to the qubit space only, and does not indicate field 
quantization. 

In Sec.~\ref{Hybrid system},  we  propose to use the quantum-mechanical nature of these macroscopic electromagnetic fields for 
a hybrid BEC-SQUID qubit system, by coupling them to the internal states of a trapped atomic gas. In the next section, we identify the qubit basis of the BEC for such a hybrid architecture.
\section{Qubit states of the BEC}
\label{qubit BEC}
Trapped ultracold atomic BECs are realized with (composite) bosonic atoms that have integer values of total-angular-momentum $F$; this includes electronic spin, nuclear spin, and orbital contributions  \cite{BECbook}. To achieve a BEC experimentally, a large number of atoms,  approximately $10^{6}$ particles,  are prepared in the same total angular-momentum state and cooled below the BEC-transition temperature.

\begin{figure}
\includegraphics[width=0.9\columnwidth]{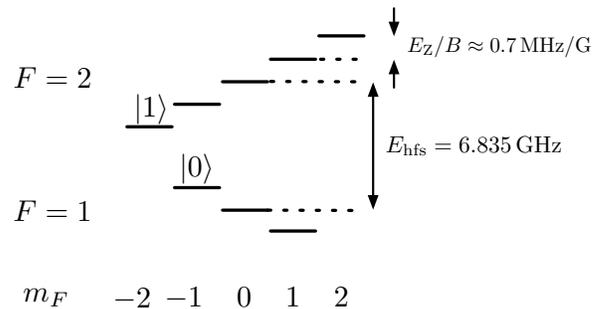}%
\caption{Energy diagram (not to scale) for the hyperfine split ground states  of $^{87}$Rb in the presence of an external Zeeman field.  The states are labeled by total angular-momentum $F$ and $z$-projection $m_{F}$. Here, the qubit basis of the BEC is formed from the states $|F=1,m^{}_{F}=-1\rangle$  and $ |F=2,m^{}_{F}=-2\rangle$. The total occupations of these two states differ by only a single atom.    \label{fig3}}
\end{figure}

Rubidium-87 is a commonly used isotope to create such condensates.   Its $F=1$ and $F=2$ degenerate non-relativistic ground state is split by the hyperfine interaction, by an energy $E_{\rm hfs}\simeq6.835\, {\rm GHz}$.  Furthermore, the remaining degenerate azimuthal states in each angular-momentum manifold can  be split by a static external Zeeman field, see Fig~\ref{fig3}.  An ensemble of $^{87}$Rb atoms in the same  angular-momentum state, such as  $|F=1,m_F=-1\rangle$, can  be used to form a  BEC. 

In the following we show how to construct qubit states using two of the BEC's internal atomic states, one of which is macroscopically populated.   In second quantization the BEC Hamiltonian is modeled by 
\begin{align}
 \label{full BEC Hamiltonian}
\hat H_{\rm B}&=\sum_{\sigma=\uparrow,\downarrow}\int d{\bm r}\,\hat\Psi^{\dagger}_{\sigma}({\bm r})\left[-\frac{\nabla^{2}}{2m}+\omega_{\sigma}+V_{}({\bm r})\right] \hat \Psi^{}_{\sigma}({\bm r})
 \nonumber\\&+\frac{1}{2}\sum_{\sigma,\sigma'}{\sf g}^{}_{\sigma,\sigma'}\int d{\bm r}\,\hat\Psi^{\dagger}_{\sigma}({\bm r})\hat\Psi^{\dagger}_{\sigma'}({\bm r})\hat\Psi^{}_{\sigma'}({\bm r})\hat\Psi^{}_{\sigma}({\bm r}),
 \end{align}
where $V_{}({\bm r})$ is the center-of-mass trapping potential and $\omega_{\sigma}$ corresponds to the internal energy of each angular-momentum state.  To form a qubit only two such angular-momentum states are needed.  For definiteness and without loss of generality we take for these two states the $^{87}$Rb states: $|\!\!\downarrow\rangle= |5^{2}S_{1/2},F=1,m^{}_{F}=-1\rangle$  and $|\!\!\uparrow\rangle= |5^{2}S_{1/2},F=2,m^{}_{F}=-2\rangle$, which from now on are  labeled by a pseudospin index $\sigma=(\downarrow,\uparrow)$. These states, in the presence of a small Zeeman field,  are separated in energy by $\omega_{\uparrow}-\omega_{\downarrow}\approx E_{\rm hfs}$, see Fig.~\ref{fig3}.    The interaction constants ${\sf g}^{}_{\sigma,\sigma'}$ correspond to the strengths of short-ranged pseudo-potentials used to model the complicated spin dependent atom-atom interactions and are, in the Born approximation,  proportional to $s$-wave scattering lengths.  

We now derive an approximation to the full Hamiltonian that captures the effects of coupling a 
flux qubit to an atomic BEC on the mean-field level. As we will see in the next section, the magnetic interaction between the BEC and SQUID only couples  states in the BEC Hilbert space that differ by a single pseudospin. Thus, in the weak coupling limit, this leads us to conclude that the full Hilbert space of the BEC can be approximated by a two-dimensional subspace: the ``ground state'' with $N$ spin-$\downarrow$ and zero spin-$\uparrow$ atoms and an ``excited state''  with $N-1$ spin-$\downarrow$ and a single spin-$\uparrow$.  This is analogous to the so-called rotating wave approximation, where the dominant process is the periodic transfer of a single quantum of energy from one subsystem to the other, i.e., the Rabi cycle. 

At zero temperature the two many-body Fock states described above, differing by a single pseudospin, are the exact ground states of \eqref{full BEC Hamiltonian}, which fully  describe the condensate and non-condensed atoms, with slightly different densities. In principle, to proceed one would represent \eqref{full BEC Hamiltonian} in this two-dimensional basis. But as the exact grounds states are unknown, except possibly numerically for a small number of atoms, one has to perform a physically motivated approximation that captures the dominant effects. As the macroscopically occupied  mode of the BEC  makes the major contribution to any matrix element involved, we approximate the true ground states by Fock states with all the atoms in the BEC,
\begin{align}
\label{interacting two-level basis}
|0\rangle^{}_{\rm B}=&\frac{1}{\sqrt{N!}}\big(\hat a^{\dagger}_{\downarrow}\big)^{N}|{\rm vac}\rangle \nonumber\\
|1\rangle^{}_{\rm B}=&\frac{1}{\sqrt{(N-1)!}}\big(\hat a^{\dagger}_{\downarrow}\big)^{N-1}\hat a^{\dagger}_{\uparrow}|{\rm vac}\rangle. \end{align}
The above approximation of course neglects number fluctuations of the BEC and non-condensed atoms, but captures the 
leading order dependence on the condensate numbers of matrix elements of the Hamiltonian .  The  spatial wave function $\phi^{}_{\sigma}({\bm r})$ of the single-particle state that corresponds to the creation and annihilation operators $a^{\dagger}_{\sigma}$ and $a^{}_{\sigma}$ can be found within mean-field theory, i.e., from a Gross-Pitaevski\v\i\, solution. 

As the BEC Hamiltonian \eqref{full BEC Hamiltonian} conserves spin, its off-diagonal elements in the basis of \eqref{interacting two-level basis} vanish, while the diagonal ones give the mean-field energies $E^{}_{0,N}, E^{}_{1,N-1}$;
\begin{align}
\label{mean field qubit BEC Hamiltonian}
\hat{H}_{\rm B}&\approx \left(\begin{array}{cc}E^{}_{1,N-1} & 0 \\0 & E^{}_{0,N}\end{array}\right)\nonumber\\&
=\frac{E^{}_{1,N-1}+E^{}_{0,N}}{2}\openone+\frac{E^{}_{1,N-1}-E^{}_{0,N}}{2}\sigma_{z}.
\end{align}
Using that $E^{}_{1,N-1}-E^{}_{0,N}\simeq \omega^{}_{\uparrow}-\omega^{}_{\downarrow}\approx E_{\rm hfs}$,  
(the interaction renormalization of  the hyperfine splitting being small compared to the level spacing) and omitting constant terms, the Hamiltonian \eqref{mean field qubit BEC Hamiltonian} in the qubit basis, can finally be written in the simple form
\begin{equation}
\label{qubit BEC Hamiltonian}
\hat{H}_{\rm B}=\frac{E_{\rm hfs}}{2}\sigma_{z}.
\end{equation} 
\section{BEC-SQUID Hybrid system}
\label{Hybrid system}
\begin{figure}
\includegraphics[width=0.75\columnwidth]{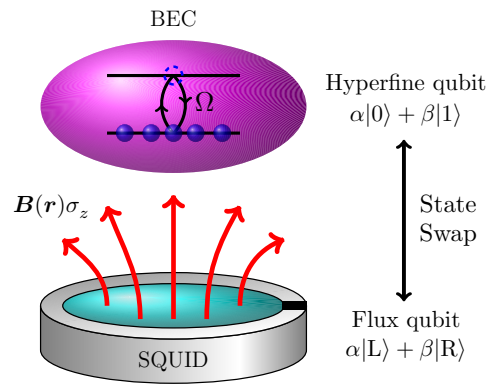}%
\caption{\label{fig4} A schematic representation of a BEC-SQUID hybrid (not to scale). 
The flux qubit is created from energy eigenstates of the SQUID, formed by superposed left- and right-going currents, and coupled by its magnetic field to the hyperfine transition of an atomic BEC, where the Rabi process periodically flips one atom from a given hyperfine spin state to the other.}
\end{figure}

By coupling the BEC-SQUID together, a hybrid  system can be formed. The coupling is done by using the electromagnetic field of the flux qubit to induce transitions between the spin states of the BEC and conversely atomic transitions in the BEC can energetically excite the flux states of the SQUID, see Fig.~\ref{fig4}.  
As the hyperfine states of the BEC considered here have the same spatial symmetry, which forbids electric dipole transitions, the transition is dominated by magnetic dipole coupling.  The magnetic coupling of the BEC to the SQUID is given by the canonical expression   
\begin{equation}
\label{BEC-SQUID coupling term}
\hat H_{\rm int}=-\sum_{\sigma,\sigma'}\int d{\bm r}\,\hat\Psi^{\dagger}_{\sigma}({\bm r})\big[{\bm \mu}\big]_{\sigma,\sigma'}\hat\Psi_{\sigma'}({\bm r})\otimes \hat{{\bm B}}({\bm r}), 
\end{equation}
multiplying the magnetic moment density of an atom with the local magnetic field and integrating over space.
Here, $\bm \mu$ is the total magnetic moment matrix of atoms in the $\sigma$ basis.  In the qubit basis of the BEC \eqref{interacting two-level basis} and using Eq.~\eqref{magnetic field operator} for the magnetic field operator of the SQUID
\begin{align}
\label{Full BEC-SQUID interaction in BEC qubit basis}
\hat H_{\rm int}&\approx\nonumber\\&-\left(\begin{array}{cc}(N-1){\bm g}_{\downarrow,\downarrow}\cdot\boldsymbol{\mu}_{\downarrow,\downarrow}+{\bm g}_{\uparrow,\uparrow}\cdot\boldsymbol{\mu}_{\uparrow,\uparrow} & \sqrt{N}{\bm g}_{\downarrow,\uparrow}\cdot\boldsymbol{\mu}_{\downarrow,\uparrow} \\\sqrt{N}{\bm g}^{}_{\uparrow,\downarrow}\cdot\boldsymbol{\mu}_{\uparrow,\downarrow} & N{\bm g}_{\downarrow,\downarrow}\cdot\boldsymbol{\mu}_{\downarrow,\downarrow}\end{array}\right)\nonumber\\&\hspace{0.4cm}\otimes \sigma_{z},
\end{align}
where  ${\bm g}^{}_{\sigma,\sigma'}=\int d{\bm r}\, \phi^{*}_{\sigma}({\bm r}) {\bm B}({\bm r})\phi^{}_{\sigma'}({\bm r})$. The BEC provides a bosonic enhancement of the single-particle matrix elements, leading to the factors of $N$ and $\sqrt{N}$.  Introducing a complex Rabi frequency $\Omega=\sqrt{N}{\bm g}_{\downarrow,\uparrow}\cdot\boldsymbol{\mu}_{\downarrow,\uparrow} $ Eq.~\eqref{Full BEC-SQUID interaction in BEC qubit basis} can be written as
\begin{align}
\label{Full BEC-SQUID interaction in BEC qubit basis part two}
\hat H_{\rm int}&=-\frac{(2N-1){\bm g}_{\downarrow,\downarrow}\cdot\boldsymbol{\mu}_{\downarrow,\downarrow}+{\bm g}_{\uparrow,\uparrow}\cdot\boldsymbol{\mu}_{\uparrow,\uparrow}}{2}\openone\otimes\sigma_{z}\nonumber\\&-\frac{{\bm g}_{\uparrow,\uparrow}\cdot\boldsymbol{\mu}_{\uparrow,\uparrow}-{\bm g}_{\downarrow,\downarrow}\cdot\boldsymbol{\mu}_{\downarrow,\downarrow}}{2}\sigma_{z}\otimes\sigma_{z}-\left(\begin{array}{cc}0& \Omega  \\\Omega^{*} & 0\end{array}\right)\otimes \sigma_{z}.
\end{align}
The first term of \eqref{Full BEC-SQUID interaction in BEC qubit basis part two} simply leads to a renormalization of the SQUID states and can be absorbed by a rescaling of the flux qubit parameters, while the $\sigma_{z}\otimes\sigma_{z}$ interaction is much smaller than the last term and can be neglected in comparison.  Thus, the qubit-qubit coupling term is taken to be
\begin{equation}
\label{qubit-qubit coupling}
\hat H_{\rm int}\approx-\left(\begin{array}{cc}0& \Omega  \\\Omega^{*} & 0\end{array}\right)\otimes \sigma_{z}. 
\end{equation}

An estimate for the Rabi frequency can be obtained by calculating the magnetic field of the SQUID from \eqref{closed form for the vector potential} using a SQUID radius of $d=1\, \mu{\rm m}$  that is carrying a current of $I=1$\, mA.  The BEC spatial wave functions $\phi_{\sigma}({\bm r})$ can be roughly approximated by the ground state of a 3-D harmonic oscillator, with a trapping frequency of $\omega_{\rm ho}=2\pi \times 50$\, Hz.  Then, for a condensate with $N=10^{6}$ atoms and a center-to-center BEC-SQUID separation of  $50\, \mu{\rm m}$,  $|\Omega|\approx 0.1$ {\rm MHz}.  A larger coupling can be achieved by moving the BEC closer to the SQUID. For example, for a BEC-SQUID separation of only  $10\, \mu{\rm m}$ leads to  $|\Omega|\approx 10$ {\rm MHz}.  There are, however, potential detrimental effects caused by  having the BEC in too  close proximity 
to a (comparatively hot) surface, such as increased heating or spin flips caused by thermal emission from the SQUID \cite{Cano,Kasch}.  

Finally, putting together the Hamiltonians of the flux qubit \eqref{flux qubit Hamiltonian}, the hyperfine qubit \eqref{qubit BEC Hamiltonian}, and coupling \eqref{qubit-qubit coupling}, the total Hamiltonian of the hybrid system is
\begin{equation}
\label{qubit-qubit Hamiltonian}
\hat{H}=\hat{H}_{\rm B}\oplus\hat{H}_{\rm S}+\hat H_{\rm int},
\end{equation}
where the direct sum is defined by $\hat{A}\oplus\hat{B}=\hat{A}\otimes\openone+\openone\otimes\hat{B}$.  Next, we show how to manipulate quantum information within this hybrid architecture. 
\section{State transfer \& BEC-SQUID entanglement}
\label{state transfer fidelity}
Here, we show that one can transfer arbitrary qubit states initially  prepared in the SQUID to the BEC, with high fidelity.   This is done by dynamically manipulating the energy level spacing of the  flux qubit, bringing the two subsystems into and out of resonance for half a Rabi period.  Although the energy level spacing of flux qubits have been experimentally tuned on sub-nanosecond time scales \cite{Paauw,Castellano},  to remain in the low-energy qubit subspace of the SQUID, the level modulation should be quasi-adiabatic. This requires the inverse of the time scale of the dynamics to be much smaller than the energy needed to leave the flux qubit subspace, which is typically on the order of several GHz.   To model this, we allow the tunneling parameter appearing in the flux qubit Hamiltonian \eqref{flux qubit Hamiltonian} to become time dependent, i.e., $\Delta\to \Delta(t)$. Furthermore, for simplicity we assume the left and right current states are degenerate, i.e.,  setting $\varepsilon=0$ in \eqref{flux qubit Hamiltonian}.  In the {\it energy eigenstate} basis the flux qubit Hamiltonian is then
\begin{equation}
\hat{H}_{\rm S}=\frac{\Delta(t)}{2}\sigma_{z},
\end{equation} 
with corresponding energy eigenstates 
\begin{align}
\label{SQUID energy basis}
|0\rangle_{\rm S}&=\frac{1}{\sqrt{2}}\big(|{\rm L}\rangle+|{\rm R}\rangle)\nonumber\\
|1\rangle_{\rm S}&=\frac{1}{\sqrt{2}}\big(|{\rm L}\rangle-|{\rm R}\rangle).
\end{align}
Along with the hyperfine qubit basis \eqref{interacting two-level basis}, the computational basis is taken to be $|ij\rangle=|i\rangle_{\rm B}\otimes|j\rangle_{\rm S}$.  The total Hamiltonian \eqref{qubit-qubit Hamiltonian} in this basis is then
\begin{align}
\label{qubit-qubit Hamiltonian in the computational basis}
\hat{H}(t)=\frac{E_{\rm hfs}}{2}\sigma_{z}\oplus\frac{\Delta(t)}{2}\sigma_{z}-\left(\begin{array}{cc}0& \Omega  \\\Omega^{*} & 0\end{array}\right)\otimes \sigma_{x}.
\end{align}

To transfer an arbitrary  qubit state that is first prepared in the flux qubit to the BEC, the general time-dependence of $\Delta(t)$ should be as follows: Initially the two subsystems should be out of resonance, $E_{\rm hfs}\neq\Delta(t)$; then, over a short time period, the ramp time, the two systems are brought into resonance, $E_{\rm hfs}=\Delta(t)$, for half a Rabi period $T_{\rm R/2}=\pi/(2|\Omega|)$;  finally, the two qubits are brought out of resonance again, leaving the initial flux qubit state in the hyperfine qubit with some final fidelity.  Specifically we set $\Delta(t)=E_{\rm hfs}W(t)$, where $W(t)$ is a unitless function of time, that is composed of hyperbolic tangents that smoothly ramp the system into and out of resonance \cite{Window function}.  The ramping time is defined as the time over which the level spacing changes from being  1\% greater than its off resonance value,  chosen to be  $E_{\rm hfs}/2$, to  within 1\% of being on resonance.

\begin{figure}
\includegraphics[width=\columnwidth]{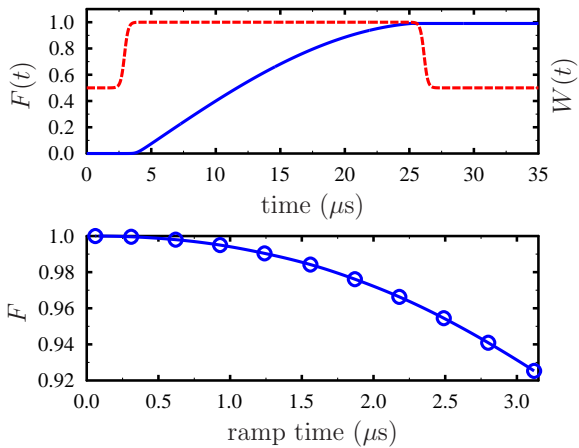}%
\caption{\label{fig5} The top panel shows the time-dependent fidelity of the direct transfer $|01\rangle\to |10\rangle$ from the flux qubit to the BEC states. The dashed line shows the profile of the function $W(t)$  used to bring the two systems into and out of resonance for half a Rabi period, with a ramping time of approximately 1 $\mu\rm s$ \cite{Window function}.  The bottom panel shows the final fidelity of the transfer as a function of ramp times.  }
\end{figure}
\begin{figure}
\includegraphics[width=\columnwidth]{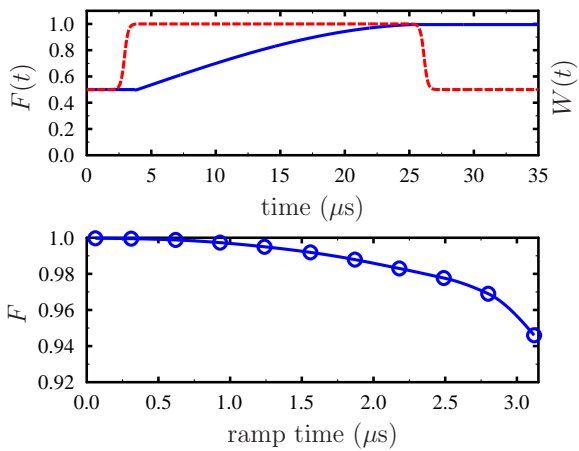}%
\caption{\label{fig6}The top panel shows the time-dependent fidelity of the transfer of the superposition state  $2^{-1/2}\big(|00\rangle+i|01\rangle\big)\to 2^{-1/2}\big(|00\rangle+i|10\rangle\big)$. As in Fig.~\ref{fig5}, the dashed line shows the profile of the function $W(t)$  used to bring the two systems into and out of resonance, with a ramping time of approximately 1 $\mu\rm s$ \cite{Window function}.  The bottom panel shows the final fidelity of the transfer as a function of ramp times. }
\end{figure}

The fidelity of the transfer is defined as $F(t)=|\langle \Psi_{\rm target}|\Psi(t)\rangle|$, where $|\Psi(t)\rangle$ is the time evolved initial state, and $|\Psi_{\rm target}\rangle$ is the sought-after  final  state. The time evolution of $|\Psi(t)\rangle$ is numerically determined using the time-dependent Hamiltonian \eqref{qubit-qubit Hamiltonian in the computational basis}.   Figure \ref{fig5} shows the fidelity during the transfer of the state $|\Psi\rangle=|01\rangle$ and the dependence of the final fidelity on the ramping time. Figure~\ref{fig6} shows the same for the superposition state $|\Psi\rangle=2^{-1/2}\big(|00\rangle+i|01\rangle\big)$.  As can be seen from Figs. \ref{fig5} and \ref{fig6} high fidelities, greater than 99\%, can  be achieved, even for relatively long ramping times, and, for a given ramping time, the final transfer fidelity is weakly state dependent.   The state can be transferred back to the SQUID by again bringing the two system into and out of resonance.  Therefore, with the exceptionally long coherence times of the hyperfine states, this system is an ideal prospective candidate for long-term storage of quantum information. 

\begin{figure}
\includegraphics[width=\columnwidth]{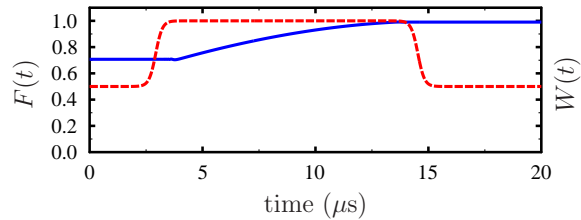}%
\caption{\label{fig7} Here the fidelity of producing  the maximally entangled qubit-qubit state $|\Psi\rangle=2^{-1/2}\big(|01\rangle+|10\rangle\big)$ is shown. This is done by first preparing the system in the state $|01\rangle$ and then bringing the two subsystems into resonance for only a quarter of a Rabi period. The ramp time used, approximately 1 $\mu\rm s$,  is the same as in Figs.~\ref{fig5} and \ref{fig6}.}
\end{figure}

Additionally, instead of transferring a state from one subsystem to the other, one can entangle the BEC and SQUID.  This is done by first preparing the system in $|\Psi\rangle=|01\rangle$ and then bringing the two qubits into and out of resonance for only a quarter of a Rabi period.  This will produce the  maximally entangled state $|\Psi\rangle=2^{-1/2}\big(|01\rangle+|10\rangle\big)$, see Fig.~\ref{fig7}. For such states, quantum correlations between the two subsystems can be probed, for example, by Bell's inequalities.  In the next section we describe how to experimentally determine the hyperfine qubit state, which is needed to test such relations,  using single-atom  spectroscopy.
\section{Tomography of the BEC qubit}
\label{BEC tomography}
\begin{figure}
\includegraphics[width=0.75\columnwidth]{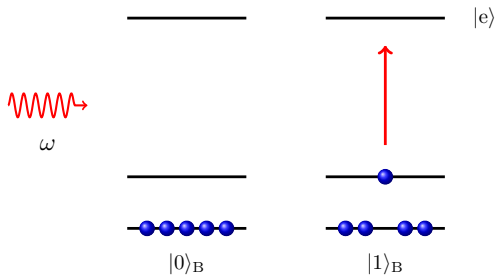}%
\caption{To measure the $\hat S_{z}$ operator component of the BEC pseudospin, an external laser source can be used to probe the occupation of the upper hyperfine state.  If there is no absorption for light tuned to the transition between the upper hyperfine state and some experimentally convenient higher excited state $|{\rm e}\rangle$, e.g. another empty hyperfine level,   then this would correspond to the system being in the $|0\rangle_{\rm B}$, or spin-$\downarrow$, state. On the other hand, if absorption occurs, this would correspond to the $|1\rangle_{\rm B}$ qubit state.  \label{fig8}}
\end{figure}

Measuring the state of the hyperfine qubit has to be delicately handled.  In principle one could independently measure the density of each pseudospin. The existence of a nonzero spin-$\uparrow$ density would indicate the $|1\rangle_{\rm B}$ state.  But this relies on determination of densities on the single atom level, a possible but quite demanding task, that also destroys the trapped BEC.   Instead we propose a nondestructive scheme using spectroscopy to determine the occupation of the spin-$\uparrow$ hyperfine state.    By measuring the absorption of an external radiation source tuned to the transition energy of the spin-$\uparrow$ state and an unoccupied level,  one effectively performs a single measurement of the hyperfine pseudospin  $\hat{S}_{z}=\frac{1}{2}\sigma_{z}$ operator,   see Fig.~\ref{fig8}.  Such a measurement would collapse the BEC qubit onto either the $|0\rangle_{\rm B}$ or $|1\rangle_{\rm B}$ state and could be used to probe entangled states of the BEC-SQUID system.  

To fully characterize the  state of the BEC qubit requires reconstructing its density matrix from measurements of a complete set of observables, so-called quantum tomography \cite{QCbook}.  A general density matrix of a pseudospin-$\frac{1}{2}$ system can be written as
\begin{equation}
\rho=\frac{1}{2}\big(\openone+{\bm a}\cdot\boldsymbol{\sigma}\big),
\end{equation}
where ${\bm a}\in \mathbb{R}$ such that $\|\bm a\|=1$, and $\boldsymbol{\sigma}=(\sigma_{x},\sigma_{y},\sigma_{z})$.  Using the facts that ${\rm Tr}\, \sigma_{i}=0$ and ${\rm Tr}\, \sigma_{i}\sigma_{j}=2\delta_{i,j}$ one can show
\begin{subequations}
\begin{align}
\langle \hat{S}^{}_{x}\rangle={\rm Tr}\,\rho \hat{S}^{}_{x}&=a_{x}/2\\ \langle \hat{S}^{}_{y}\rangle={\rm Tr}\,\rho \hat{S}^{}_{y}&=a_{y}/2\\ \langle \hat{S}^{}_{z}\rangle={\rm Tr}\,\rho \hat{S}^{}_{z}&=a_{z}/2,
\end{align}
\end{subequations}
where $\hat{S}_{i}=\frac{1}{2}\sigma_{i}$  are the spin operators.  Thus, by experimentally measuring the expectation values for each spin direction the density matrix can  be reconstructed.  

As described above, the expectation value of $\hat{S}^{}_{z}$ can be found by averaging  repeated measurements of identically prepared states of the BEC qubit, i.e., states prepared in the flux qubit, passed to the BEC, and then measured.   As only $\hat{S}^{}_{z}$ is directly measurable a modified protocol must be used to determine the expectation value of  $\hat{S}^{}_{x}$ and $\hat{S}^{}_{y}$. To obtain the expectation value of  $\hat{S}^{}_{x}$ and $\hat{S}^{}_{y}$ for a given state or density matrix $\rho$ one can measure $\hat{S}^{}_{z}$ of a rotated state $\rho'$.   To see this one notes that $S_{z}$ is related to  $S_{x}$ and $S_{y}$ by a unitary transformation; 
\begin{subequations}
\begin{align}
\hat{S}^{}_{x}&=e^{-i\frac{\pi}{2}\hat{S}^{}_{y}}\hat{S}^{}_{z}e^{i\frac{\pi}{2}\hat{S}^{}_{y}}\\\hat{S}^{}_{y}&=e^{i\frac{\pi}{2}\hat{S}^{}_{x}}\hat{S}^{}_{z}e^{-i\frac{\pi}{2}\hat{S}^{}_{x}}.
\end{align}
\end{subequations}
Thus to obtain $\langle \hat{S}^{}_{x}\rangle$ with respect to $\rho$ 
\begin{align}
\langle \hat{S}^{}_{x}\rangle&={\rm Tr}\,\rho e^{-i\frac{\pi}{2}\hat{S}^{}_{y}}\hat{S}^{}_{z}e^{i\frac{\pi}{2}\hat{S}^{}_{y}}={\rm Tr}\,e^{i\frac{\pi}{2}\hat{S}^{}_{y}}\rho e^{-i\frac{\pi}{2}\hat{S}^{}_{y}}\hat{S}^{}_{z}\nonumber\\&={\rm Tr}\,\rho' \hat{S}^{}_{z},
\end{align}
where $\rho'=e^{i\frac{\pi}{2}\hat{S}^{}_{y}}\rho e^{-i\frac{\pi}{2}\hat{S}^{}_{y}}$.  Hence, one prepares and measures $\hat{S}^{}_{z}$ for the rotated state $\rho'$ to obtain $\langle \hat{S}^{}_{x}\rangle$ for the desired state $\rho$ and similarly for $\langle \hat{S}^{}_{y}\rangle$.  Using these values to reconstruct the desired density matrix, the fidelity of the state transfer can be experimentally determined. 

\section{Conclusion}
We have proposed a novel hybrid quantum  system, which could be an ideal candidate for long-term storage of qubit states, comprised of the hyperfine states of an atomic BEC magnetically coupled to a flux qubit.  The potential of almost infinite coherence times of the hyperfine qubit is advantageous and highly desirable, while the qubit-qubit coupling, and thus the transfer time, can be  tuned by adjusting the BEC-SQUID separation.  Furthermore, we propose a straightforward method of using atomic spectroscopy to experimentally determine the hyperfine qubit density matrix, which would enable the determination of the state transfer fidelity or quantum correlations of entangled BEC-SQUID systems.   

Additional open questions that are left for future work concern the effects of interactions and beyond mean-field effects, as well as finite temperature on the BEC state, both of which lead to decoherence and  depletion of  the condensate state.   These are potentially difficult effects to capture accurately, as the simple two-dimensional qubit Hilbert space has to be enlarged to infinite dimensions. 

\begin{acknowledgments}
This research was supported by the NRF of Korea, grant Nos. 2010-0013103 and 2011-0029541. 
URF thanks D. Cano, J. Fort\'agh, R. Kleiner, N. Schopohl, and C. Zimmermann for helpful discussions.
\end{acknowledgments}


\begin{thebibliography}{99}
\bibitem{Window function} Explicitly, $W(t)=\Theta(\kappa,t)\Theta(\kappa,\tau-t)+E^{}_{\rm or}\Theta(\kappa,-t)+E^{}_{\rm or}\Theta(\kappa,t-\tau)$, 
where $\Theta(\kappa,t)=\frac{1}{2}+\frac{1}{2}\tanh(\kappa t)$, $\tau$ is the time the two systems are on resonance (typically for a half or a quarter of a Rabi cycle), and  $E^{}_{\rm or}$ is the off-resonance value of the SQUID tunneling amplitude, $\Delta$, in units of $E_{\rm hfs}$ (we set $E^{}_{\rm or}=\frac{1}{2}$ for all our calculations).  
The function $\Theta(\kappa,t)$ is a smoothed step function, such that $\lim_{\kappa\rightarrow\infty}\Theta(\kappa,t)=\Theta(t)$; different
ramp times are obtained by varying  $\kappa$.
\end{thebibliography}
\end{document}